%% file: main.tex
\documentclass[trackchanges,twocolumn]{aastex7}
\usepackage{xspace}

\usepackage{graphicx,caption,subcaption}
\usepackage[dvipsnames]{xcolor}

\newcommand{\Msun}{M$_{\odot}$\xspace}

\newcommand{\Msunyr}{M$_{\odot}$ yr$^{-1}$\xspace}
\newcommand{\Msunyrkpc}{M$_{\odot}$ yr$^{-1}$ kpc$^{-2}$\xspace}
\newcommand{\sigmasfr}{$\Sigma_{\mathrm{SFR}}$\xspace}

\newcommand{\ha}{H$\alpha$\xspace}

\newcommand{\um}{$\mu$m }
\newcommand{\gaia}{\textit{Gaia}\xspace}
\newcommand{\spitzer}
{\textit{Spitzer}\xspace}

\newcommand{\hst}
{\textit{HST}\xspace}
\newcommand{\jwst}
{\textit{JWST}\xspace}
\newcommand\dolphot{\texttt{DOLPHOT}\xspace}

\newcommand\comment[1]{$^{#1}$}
\newcommand\mum{$\mu$m\xspace}
\newcommand\ikcmd{F814W$-$F200W\xspace}
\newcommand\ikcmdnospace{F814W$-$F200W}

\begin{document}

\title{Evolved Supergiants in PHANGS I: Red Supergiants in 19 Galaxies between 5-20 Mpc with HST and JWST}
\input{authors}
\correspondingauthor{Sumit K. Sarbadhicary}

\begin{abstract}
Red supergiants (RSGs) are important for our understanding of supernova progenitors, stellar populations, stellar evolution, mass loss and dust production. Extragalactic surveys of RSGs have a long history in the Local Group, but few studies exist beyond that due to the limited resolution and sensitivity of ground-based and previous space-based infrared observatories. Here we demonstrate the combined power of \hst and \jwst to push systematic searches of RSGs out to $\sim$20 Mpc. We introduce a catalog of 97057 RSGs ---the largest single-survey release of RSGs--- with masses $\gtrsim$10 \Msun in 19 galaxies from the PHANGS HST+JWST Treasury program. We use \hst F814W and \jwst F200W photometry to select stars as RSGs based on predicted colors and magnitudes from PARSEC isochrones. The spatial distribution of our recovered RSGs follow the familiar pattern of mostly being concentrated in active star-forming regions such as spiral arms and central starburst rings. The RSG number density on kpc-scales is strongly correlated ($r_s$$\sim$0.82) with local star-formation rate density ($\Sigma_{SFR}$) traced by extinction-corrected far-ultraviolet (FUV) from \emph{GALEX+WISE}, and weakly correlated ($r_s$$\sim$0.57) with the total stellar mass density ($\Sigma_*$), traced by near-infrared emission from \emph{WISE+Spitzer}. The number of RSGs per mass of stellar populations with ages 6-30 Myr (the likely age range of RSGs $>$10\Msun) is $\sim$1 per 10$^{3.77\pm0.27}$ \Msun, assuming constant star-formation rates from FUV+W4. Our sample will be a useful resource for tracking progenitors and feedback sites of future supernovae in PHANGS,age-dating stellar populations, and more. 
\end{abstract}
\keywords{}

\section{Introduction} 

Red supergiants (RSGs) typically refer to K or M-type supergiants that have evolved from stars with zero-age main-sequence (ZAMS) masses $>$ 8 \Msun \citep{Levesque2018}. Studies of RSGs are essential for many areas of astrophysics. They have been confirmed as progenitors of various subtypes of Type II supernovae (SNe) \citep{Smartt2009,Smartt2015}, making the understanding of their pre-SN evolution vital. In stellar evolution, RSGs are important laboratories for understanding mass-loss mechanisms in cool supergiants \citep[e.g.,][]{Maeder1981,Chiosi1986,vanLoon2005,Beasor2020}, binary evolution models \citep[e.g.,][]{Neugent2019,Patrick2020,Neugent2020, Patrick2022}, metallicity, stellar rotation, and binary fraction diagnostics \citep{Maeder1980,Massey2003,Eldridge2017,Massey2021b}. For stellar feedback studies, RSGs serve as markers of future Type II SNe in nearby galaxies, enabling spatially-resolved studies of their host stellar populations \citep[e.g.,][]{Anderson2008, Kangas2017} and future SN feedback sites \citep{MC23,Sarbadhicary2023,MC24}. 

Many of these topics have greatly benefited from extragalactic surveys of RSGs. Such studies were pioneered using ground-based photometric surveys of Local Group galaxies \citep[$\lesssim1.2$ Mpc, e.g.][]{Massey1998,Massey2002,Massey2003,Massey2009,Massey2016}. Follow-up spectroscopy confirmed many of these photometrically-identified massive stars as cool supergiants with bolometric luminosities and temperatures in the region expected for RSGs \citep[e.g.,][]{MasseyOlsen2003,Levesque2006,Drout2012, Neugent2012}. Availability of NIR photometry (e.g., \textit{JHK}) and \gaia astrometry significantly improved the fidelity of RSG catalogs by effectively removing foreground dwarfs and asymptotic giant branch (AGB) stars \citep[e.g.][]{Bonanos2009,Bonanos2010,Boyer2011,Yang2019,Neugent2020,Massey2021}. 

Surveys of RSGs in galaxies beyond the Local Group have been attempted out to distances of $\sim$5 Mpc \citep[e.g.,][]{Gazak2015,Patrick2017,Khan2017,Chun2017,Bonanos2025, Maravelias2025}. The limited resolution of ground-based optical and near-infrared observatories, or even previous-generation space observatories like \spitzer, have made identifying RSGs challenging at larger distances.

The superior angular resolution of HST and JWST allows identification of RSGs in galaxies out to $\gtrsim 20$~Mpc. This capability has been demonstrated in identification of the progenitors of individual core-collapse SNe \citep[for a review, see][]{Smartt2015,VanDyk2025} and surveys of individual galaxies \citep[e.g.,][targeting NGC 6946 and I Zw 18]{Johnson2023,Hirschhauer2024} but not previously deployed over a large sample of galaxies.

Here we push the limits of extragalactic RSG surveys and construct catalogs in 19 star-forming galaxies at distances between 5--20 Mpc using the PHANGS (Physics at High Angular Resolution in Nearby Galaxies) HST \citep{Lee2022} and JWST Treasury \citep{Lee2023} data. 
The PHANGS surveys provide substantial data for exploring the correlation between RSGs and the multi-age stellar populations and the multiphase ISM in the surrounding galaxy. Relevant catalogs in PHANGS include: molecular clouds \citep{Rosolowsky2021,Sun2022}, infant dust-embedded star clusters \citep{Rodriguez2025, Graham2025, hassani2025,Whitmore2023}, optically visible star clusters and associations \citep{Maschmann2024, Thilker2025, Larson2023}, supernova remnants \citep{Li2024}, HII regions \citep{Emsellem2022,Santoro2022,Groves2023}, and dust maps tracing the cold ISM \citep{Leroy2023, Chown2025}.

The paper is structured as follows: Section \ref{sec:obs} details the HST+JWST observations used in this paper, and supporting multiwavelength data to characterize the RSGs. Section \ref{sec:rsgselection} details the photometry, identification and spatial properties of the recovered RSGs. Section \ref{sec:results:sfrs} discusses the correlation with local kpc-averaged star-formation rates, and Section \ref{sec:conclusion} provides a summary of our results.

\begin{deluxetable*}{lrrrrrrrr}
\tablecaption{Properties of the 19 PHANGS-HST and JWST galaxies where we search for RSGs. The last two columns report the number of RSGs, $N_{\rm RSG}$, and number of younger RSGs, $N_{\rm RSG}^{\rm young}$, within the thick and thin dashed polygon regions defined in Figure \ref{fig:rsgselection}.}
\tablehead{\colhead{Name} & \colhead{RA} & \colhead{Dec} & \colhead{Distance} & \colhead{Log M$_*$} & \colhead{SFR} & \colhead{Type} & \colhead{$N_{\mathrm{RSG}}$} & \colhead{$N^{\mathrm{young}}_{\mathrm{RSG}}$}\\
\colhead{} & \colhead{} & \colhead{} & \colhead{(Mpc)} & \colhead{(M$_{\odot}$)} & \colhead{(M$_{\odot}$ yr$^{-1}$)} & \colhead{} & \colhead{} & \colhead{}}
\startdata
NGC5068 & 13:18:54.74 & -21:02:19.48 & 5.2 & 9.25 & 0.20 & Sc & 955 & 153 \\
IC5332 & 23:34:27.49 & -36:06:03.89 & 9.0 & 9.10 & 0.11 & SABc & 453 & 60 \\
NGC628 & 01:36:41.73 & +15:47:01.11 & 9.8 & 10.07 & 0.93 & Sc & 2673 & 138 \\
NGC3351 & 10:43:57.76 & +11:42:13.21 & 10.0 & 10.18 & 0.87 & Sb & 1005 & 48 \\
NGC3627 & 11:20:15.01 & +12:59:29.40 & 11.3 & 10.77 & 3.31 & Sb & 10342 & 492 \\
NGC2835 & 09:17:52.91 & -22:21:16.84 & 12.2 & 9.66 & 0.57 & Sc & 2428 & 316 \\
NGC4254 & 12:18:49.63 & +14:24:59.08 & 13.1 & 10.38 & 2.76 & Sc & 8716 & 365 \\
NGC4321 & 12:22:54.93 & +15:49:20.29 & 15.2 & 10.58 & 2.42 & SABb & 7196 & 362 \\
NGC4535 & 12:34:20.30 & +08:11:52.70 & 15.8 & 10.29 & 1.23 & Sc & 2760 & 124 \\
NGC1087 & 02:46:25.18 & -00:29:55.38 & 15.9 & 9.88 & 1.15 & Sc & 4200 & 197 \\
NGC4303 & 12:21:54.93 & +04:28:25.48 & 17.0 & 10.43 & 4.27 & Sbc & 15917 & 1648 \\
NGC1385 & 03:37:28.56 & -24:30:04.18 & 17.2 & 9.95 & 1.97 & Sc & 4807 & 239 \\
NGC1566 & 04:20:00.38 & -54:56:16.84 & 17.7 & 10.63 & 3.18 & SABb & 9468 & 736 \\
NGC1433 & 03:42:01.49 & -47:13:18.99 & 18.6 & 10.56 & 0.57 & SBa & 1862 & 109 \\
NGC7496 & 23:09:47.29 & -43:25:40.26 & 18.7 & 9.94 & 1.99 & Sb & 1911 & 147 \\
NGC1512 & 04:03:54.14 & -43:20:55.41 & 18.8 & 10.38 & 0.59 & Sa & 2057 & 169 \\
NGC1300 & 03:19:41.00 & -19:24:40.01 & 19.0 & 10.56 & 1.02 & Sbc & 2273 & 84 \\
NGC1672 & 04:45:42.49 & -59:14:50.13 & 19.4 & 10.67 & 6.62 & Sb & 12028 & 1328 \\
NGC1365 & 03:33:36.37 & -36:08:25.45 & 19.6 & 10.87 & 12.85 & Sb & 6006 & 678
\enddata
\end{deluxetable*}\label{tab:galaxies}

\section{Observations} \label{sec:obs}
In this section, we describe the HST and JWST observations for our target galaxies (Table \ref{tab:galaxies}) from which RSGs will be selected (Section \ref{sec:hstandjwst}), as well as supporting multi-wavelength data from the PHANGS MegaTables catalog that will help characterize the RSG population (Section \ref{sec:megatables}).
\subsection{The PHANGS-HST and JWST data} \label{sec:hstandjwst}
We draw our targets from the PHANGS sample which consists of 90 galaxies that were observed with the Atacama Large Millimeter Array (ALMA) \citep{Leroy2021}. A subset of about 74 of these galaxies have been systematically observed with other telescopes, including HST \citep[38 published, see][GO-15654, GO-17502]{Lee2022, Maschmann2024} and JWST \citep[][GO-2107, GO-3707]{Lee2023, Chown2025, Egorov2025}. In this paper, we work with the first 19 of these 74 galaxies observed by the PHANGS JWST Cycle 1 Treasury Survey \citep{Lee2023} because of their high-quality astrometric alignment in all filters \citep{Williams2024}, making them ideal for our first attempt at multiband photometry and census of supergiants. 

From the NUV to the MIR, the combined HST and JWST data offer 13 wide/medium filters spanning 0.25-21 \micron. These filters include the HST F275W(\emph{NUV}), F336W(\emph{U}), F438W(\emph{B}), F555W(\emph{V}), and F814W(\emph{I}), NIRCam F200W, F300M, F335M and F360M, and MIRI F770W, F1000W, F1130W, F2100W. Narrowband HST H$\alpha$ with F657N and F658N also exists for these galaxies \citep{Chandar2025}, but we mainly use the broadband filters for identifying stars. Most of the HST data were taken with the UVIS/WFC3 instrument, with only NGC~628 and NGC~1672 relying on archival ACS data. The observations, data reduction, image fidelity, and feasibility of PSF photometry have been extensively discussed in \cite{Lee2022}, \cite{Thilker2022}, and \cite{Maschmann2024} for the PHANGS-HST data, and \cite{Lee2023} and \cite{Williams2024} for the PHANGS-JWST data, and we refer the reader to these papers for details.

\subsection{Star-formation tracers from PHANGS MegaTables} \label{sec:megatables}
PHANGS provides a number of supporting datasets which we can use to understand the statistics of the recovered RSG population. In this paper we use the PHANGS MegaTables constructed by \cite{Sun2022,Sun2023}, which estimate relevant star-formation and ISM measurements from multi-wavelength maps in hexagonal bins of $\sim$ 1.95 kpc$^2$ (the large bins are necessitated by the limiting resolution of the WISE 22~\mum maps used for attenuation correction). While these bins may smooth over finer sub-kpc scale spatial correlations between RSGs and star-forming regions, we believe the MegaTables are a good starting point as all the SFR measurements and their inter-correlations have been carefully calibrated and vetted by \cite{Sun2023}. The MegaTable apertures are large enough to avoid low number statistics at low SFRs, and to treat the RSGs and other star-formation within the apertures as approximately co-eval, even after some stellar drifting and mixing. We refer the reader to \cite{Sun2022} and \cite{Sun2023} for details on how the tables were constructed. 

The PHANGS MegaTables provide several star-formation rate (SFR) and stellar mass tracers on 1.5 kpc scales that can be useful for assessing the recovered RSG population, including: SFRs from \emph{GALEX} far-ultraviolet (FUV, 154~nm) and near-ultraviolet (NUV,231~nm) with \emph{WISE} W4 (22 \mum)-based extinction correction based on the prescriptions of \cite{Leroy2019} and \cite{Belfiore2023}, \ha-based SFRs from the ESO Wide Field Imager with W4-based extinction correction \citep{Calzetti2007, Murphy2011, Belfiore2023}, MUSE \ha-based SFRs with Balmer-decrement correction \citep{Groves2023}, and the \emph{WISE W1} (3.4\mum) + \emph{Spitzer IRAC} (3.6\mum) based stellar mass maps using a mass-to-light ratio based on \cite{Leroy2019}. 

In Section \ref{sec:results:sfrs}, we will use  the \cite{Belfiore2023}-based FUV+W4 hybrid SFRs and Balmer-decremented-corrected MUSE \ha SFRs, to assess the correlation and production rate of RSGs per \Msun, while briefly mentioning the degree of correlation with the other SFR maps.
\begin{figure*}
    \centering
    \includegraphics[width=\textwidth]{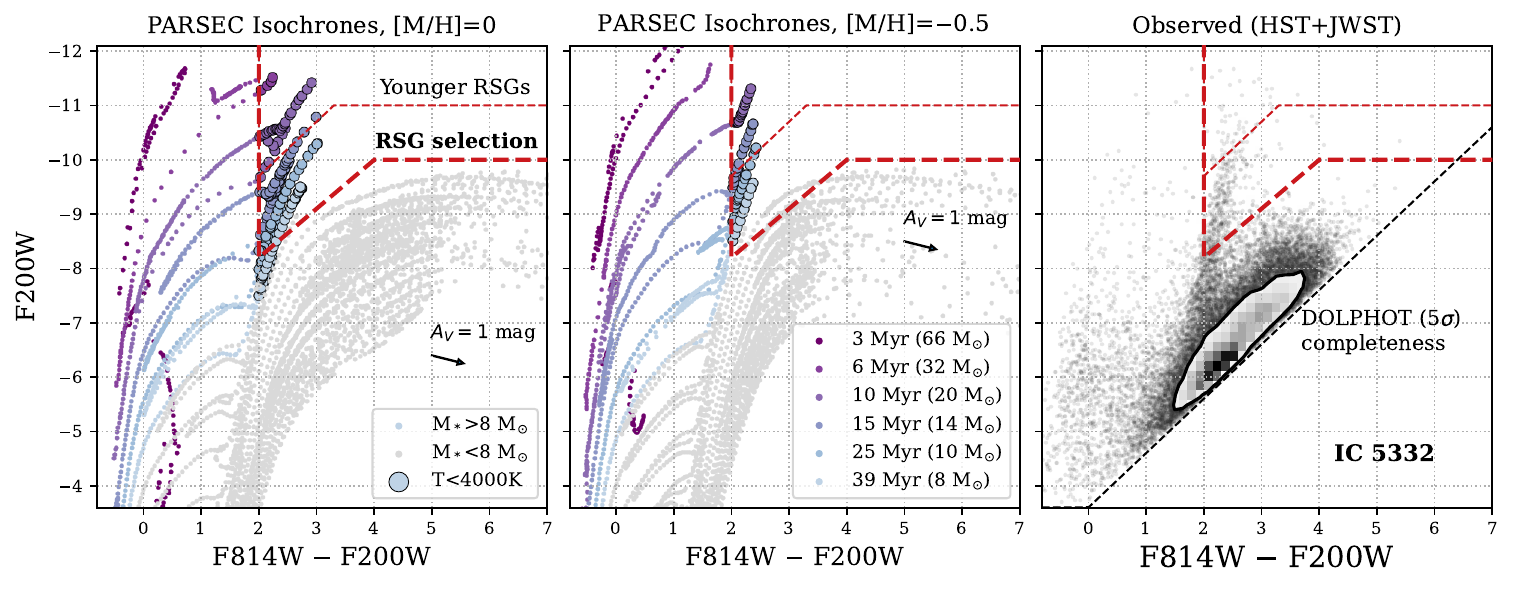}
    \caption{The selection process of RSGs from theoretical CMDs based on PARSEC isochrones. \textbf{Left+Middle:} F814W$-$F200W CMDs predicted by PARSEC for two metallicities $[M/H]=0$ and $[M/H]=-0.5$. Colored tracks are for zero-age main-sequence ages $<40$~Myr (or initial mass $>8$~\Msun), which are expected to evolve into RSGs ($\leq4000$~K). Gray tracks are for lower-mass ($<8$~\Msun) stars that may produce luminous AGBs. RSGs are selected in the region defined by the red (bold) dashed polygon, and giants with temperatures $<$4000 K in the isochrones appear as larger circles. A thinner red dashed polygon captures younger RSGs ($>$14 \Msun). The reddening vector for $A_V$=1 mag is shown with arrow, which is a typical value observed for our galaxies \textbf{Right:} Comparison of the selection box with an example \ikcmd from IC5332. The protrusion of the RSG branch on top of the C-rich AGBs are clearly visible. The black contour represents 68\% of the sample, and within this the data are plotted as 2D histograms. The black dashed line represents the 5$\sigma$ completeness cut in our DOLPHOT photometry.}    \label{fig:rsgselection}
\end{figure*}

\begin{deluxetable}{lrlr}
\tablenum{1}
\tablecaption{\dolphot parameters and their values set in this paper (Section \ref{sec:dolphot}). We refer reader to the \dolphot manual\comment{a} for the parameter descriptions.}
\tabletypesize{\scriptsize}
\tablewidth{0pt}
\tablehead{
\colhead{Parameter} & \colhead{Value} & \colhead{Parameter} & \colhead{Value} 
}
\startdata
\texttt{RAper} & \texttt{[8, 3]}\comment{*} & \texttt{RChi} & \texttt{[3.0, 1.5]}\comment{*}\\
\texttt{RSky0} & \texttt{[15, 15]}\comment{*}& \texttt{Rsky1} & \texttt{[35, 35]}\comment{*}\\
\texttt{RPSF} & \texttt{[13, 15]}\comment{*} & \texttt{aprad} & \texttt{[10, 10]}\comment{*}\\
\texttt{apsky} & \texttt{[15 25, 20 35]}\comment{*} & \texttt{RCentroid} & \texttt{2}\\
\texttt{SigFind} & 3 & \texttt{SigFindMult} & 0.85\\
\texttt{SigFinal} & 3.5 & \texttt{MaxIT} & 25\\
\texttt{FPSF} & \texttt{Lorentz} & \texttt{PSFPhot} & 1\\
\texttt{PSFPhotIt} & 3 & \texttt{FitSky} & 3\\
\texttt{SkipSky} & 1 & \texttt{SkySig} & 2.25\\
\texttt{NegSky} & 1 & \texttt{NoiseMult} & 0.1\\
\texttt{FSat} & 0.999 & \texttt{Zero} & 25\\
\texttt{PosStep} & 0.25 & \texttt{dPosMax} & 2.5\\
\texttt{RCombine} & 1.5 & \texttt{SigPSF} & 5\\
\texttt{PSFStep} & 0.25 & \texttt{MinS} & 1 \\
\texttt{MaxS} & 9 & \texttt{MaxE} & 0.5\\
\texttt{UseWCS} & 2 & \texttt{Align} & 2\\
\texttt{AlignIter} & 2 & \texttt{AlignTol} & 0\\
\texttt{AlignStep} & 1 & \texttt{AlignOnly} & 0\\
\texttt{Rotate} & 1 & \texttt{SubResRef} & 1\\
\texttt{SecondPass} & 5 & \texttt{SearchMode} & 1\\
\texttt{Force1} & 1 & \texttt{EPSF} & 1\\
\texttt{PSFsol} & 1 & \texttt{PSFres} & 1\\
\texttt{psfoff} & 0 & \texttt{ApCor} & 1\\
\texttt{SubPixel} & 1 & \texttt{ForceSameMag} & 0\\
\texttt{FlagMask} & 4 & \texttt{CombineChi} & 0\\
\texttt{WFPC2useCTE} & 1 & \texttt{ACSuseCTE} & 0\\
\texttt{WFC3useCTE} & 0 & \texttt{ACSpsfType} & 0\\
\texttt{WFC3UVISpsfType} & 0 & \texttt{WFC3IRpsfType} & 0\\
\enddata
\tablecomments{
\comment{a}\href{http://americano.dolphinsim.com/dolphot/dolphot.pdf}{DOLPHOT manual}, \cite{Dolphin2000}.\\
\comment{*}Comma-separated pair of values were used for HST (WFC3/ACS) and JWST (NIRCam) data, respectively.}
\end{deluxetable}\label{tab:dolphotparams}

\section{Red Supergiant Selection} \label{sec:rsgselection}

We use the HST and JWST data introduced in Section \ref{sec:obs} to find RSGs in our 19 targets. We use the F814W (0.8 \mum) and F200W (2 \mum) filters for the selection. These filters are close to the peak of the RSG spectral energy distribution (SED), which typically occurs around $\sim1$~\mum. This makes them the most suitable among our available filters for identifying RSGs (Figure~\ref{fig:rsgselection}). To construct the catalogs, we start with PSF photometry of genuine stars with \dolphot (Section \ref{sec:dolphot}). We select RSGs based on predicted \ikcmd color and magnitudes (Section \ref{sec:rsgidentification}). The we report the observed colors, magnitudes (Section \ref{sec:compdata}), and spatial properties (Section \ref{sec:spatialprops}) of recovered RSGs in the sample of 19 galaxies.

\subsection{PSF photometry \& point source selection with DOLPHOT} \label{sec:dolphot}
We carry out multiband PSF photometry with \dolphot, which is designed to operate on HST \citep{Dolphin2000, Dolphin2016} and JWST data \citep{Weisz2024}. PSF and pixel maps of the HST WFC3/UVIS, ACS/WFC and NIRCam were downloaded from the \dolphot website. Prior to photometry, the \texttt{nircammask}, \texttt{wfc3mask} and \texttt{acsmask} tasks were used to mask out bad or saturated pixels, and make minor adjustments to the FITS headers to aid the \dolphot routine. For each dither, a background sky map is generated with the \texttt{calcsky} package. With this setup, the \dolphot photometry routine was run on the individual dither images, with a mosaicked F814W image of each galaxy being used as reference for alignment. We chose F814W as the reference image as it is roughly the midpoint of the full NUV-to-NIR range spanned by our filters, and is therefore most likely to have enough stars in common with images taken using other filters for accurate astrometric alignment, PSF and aperture corrections. 

The routine \dolphot iteratively detects sources, records their pixel locations and aperture-corrected fluxes using the procedure in \cite{Dolphin2000} and several user-defined parameters. The key \dolphot parameters used in this paper are summarized in Table \ref{tab:dolphotparams}, mostly based on recommendations by the JWST Resolved Stellar Populations ERS program \citep{Weisz2024}. 


We then optimize our run to select point sources, which describes our RSGs as opposed to slightly extended objects like clusters \citep[see discussion in Section~4 of][]{Thilker2022}. We will use the F814W (0.8 \mum) and F200W (2 \mum) filters for the selection of RSGs from the \dolphot catalog. These filters are close to the peak of the RSG SEDs typically occurring around $\sim1$~\mum, making them the most suitable among our available filters for identifying RSGs (Figure~\ref{fig:rsgselection}). We first need to obtain a photometric sample of stars that is mostly uncompromised by crowding, and reject stars of poor or dubious photometric quality. \dolphot records quantities such as photometry signal-to-noise, crowding, roundness, and goodness-of-fit $\chi^2$ in each band \citep[see][for definitions]{Dolphin2000}, and we again follow \citet{Weisz2024} to set thresholds to select a candidate star catalog (or `good stars'):
\begin{enumerate}
\itemsep -1em
\item The detection signal-to-noise in both F814W and F200W filters should be greater than 5:\\
\texttt{S2N\_WFC3\_F814W $\geq$ 5 \& S2N\_NIRCAM\_F200W $\geq$ 5}\\
\item The \emph{crowding} parameter, which informs how much brighter a star would have been in the absence of nearby fitted stars, should be less than 0.2 mag in both F814W and F200W:\\\texttt{Crowd\_WFC3\_F814W $\leq$ 0.2 \& Crowd\_NIRCAM\_F200W $\leq$ 0.2}\\
\item The \emph{sharp}-squared parameter should be less than 0.02 in both filters. For a perfectly fit star, \emph{sharp}=0. For overly sharp sources (e.g. cosmic rays), \emph{sharp}$>$0, while for blended sources, \emph{sharp}$<$0. As per convention, we use the squared value of this parameter: \\
\texttt{Sharp\_WFC3\_F814W$^2$ $\leq$ 0.02 \& Sharp\_NIRCAM\_F200W$^2$ $\leq$ 0.02}
\end{enumerate} 

For the remainder of this paper, we will discuss fluxes in vegamag. Analysis in the subsequent sections will use stars brighter than the 5$\sigma$ completeness limit from the \dolphot good-star selection above. Uncertainties in the \dolphot photometry are $\lesssim$0.22 mag in the F814W and F200W filters. Absolute magnitudes were obtained assuming distances to galaxies from \cite{Anand2021}, and corrected for extinction in each filter due to foreground Milky Way dust using \cite{Schlafly2011}, assuming a \cite{Fitzpatrick1999} extinction law with $R_V=3.1$. 

\subsection{Identification of Red Supergiants} \label{sec:rsgidentification}

We select RSGs from the \dolphot good-star catalog 
based on the predicted locations of RSGs in \ikcmd CMDs (Figure~\ref{fig:rsgselection}) from PARSEC v1.2S stellar isochrone models \citep{Bressan2012, Tang2014}. We use single-stellar, non-rotating isochrones with ages 3-40 Myr (corresponding to stars with initial masses $>$8 \Msun), downloaded from the \texttt{CMD} web interface\footnote{\url{http://stev.oapd.inaf.it/cmd}}.  Absolute magnitudes in our PHANGS HST and JWST filters were obtained with \texttt{CMD} using the bolometric correction database YBC \citep{Chen2019, Bohlin2020}. Isochrones were derived every 0.2~dex in logarithmic age, and sampled with the initial mass function from  \cite{Kroupa2001, Kroupa2002}. These tracks are shown in shaded color in Figure \ref{fig:rsgselection}. We show them for two value of metallicities\footnote{We will be using the PARSEC metallicity notation defined on the \texttt{CMD} interface of $[M/H]=\mathrm{log}(Z/X)-\mathrm{log}(Z/X)_{\odot}$, with $(Z/X)_{\odot}=0.0207$ and $Y=0.2485 + 1.78Z$.}, [M/H]=0 and [M/H]=$-$0.5, which roughly covers the metallicity ranges observed in these 19 galaxies \citep{Groves2023,Pessa2023}.

Aside from stellar tracks $>$8 \Msun that will evolve into RSGs, we also show 1-8 \Msun tracks from \texttt{CMD} as gray isochrones in Figure \ref{fig:rsgselection} (left, middle panels) to show the expected positions of asymptotic giant branch stars (AGBs) in \ikcmd. These are associated with copious dust production, and are a common source of contamination in RSG catalogs \citep{Massey2021}. The AGB tracks were modeled in PARSEC using COLIBRI atmopsheric models \citep{Marigo2013, Rosenfield2016, Pastorelli2019, Pastorelli2020}, and the default circumstellar dust composition of 60\% Silicate and 40\% AlO$_x$ for M-stars, and 85\% Amorphous Carbon and 15\% Silicon Carbide for C-stars \citep{Groenewegen2006, Marigo2008}. PARSEC provides several other options for AGB dust composition with variations in the C, Si, and AlO$_x$ abundances, but we found that our RSG selection defined below is robust to these variations.

We define the RSG selection region as the thick dashed polygon in Figure~\ref{fig:rsgselection}. The design of the region is intended to minimize contamination from hotter yellow supergiants and cooler AGBs/super-AGBs, and can be explained as follows: 
\begin{enumerate}
    \itemsep -0.1em
    \item The blue end of the polygon is at \ikcmdnospace=2 mag, in order to restrict to stars cooler than 4000~K, appropriate for K/M spectral type expected for RSGs \citep{Levesque2005, Levesque2018}. The color boundary should also exclude any Milky Way foreground dwarfs in the direction of our galaxies, based on experiments with the TRILEGAL code \citep{girardi2005, vanhollebeke2009}.
    \item Between \ikcmdnospace=2$-$4 mag, the F200W brightness limit is drawn diagonally from $-$8.2 mag to $-$10 mag to form a wedge, which allows inclusion of the lower mass ($<$10 \Msun) RSGs with up to moderate levels of ISM extinction ($A_V \lesssim 1.5$). However, as one can tell from Figure \ref{fig:rsgselection}, the selection of these 8-10 \Msun will be incomplete as the wedge is about 0.8 mag brighter than the base of the 8 \Msun RSG branch at solar metallicity. This was however a choice made to reduce contamination with sub-solar metallicity AGBs (compare grey points in the left and middle panels, Figure \ref{fig:rsgselection}). For the purpose of this paper, we consider our RSG sample to be effectively complete above 10 \Msun.
    \item At \ikcmdnospace$\geq$4 mag, we include all stars brighter than F200W=$-10$ mag as RSGs, since AGB contamination above this brightness is expected to be minimal based on PARSEC predictions for up to [M/H]=$-$1. The box is also allowed to span redward in \ikcmd up to the 5$\sigma$ \dolphot completeness limit (black dashed line, third panel, Figure \ref{fig:rsgselection}) in order to be as complete as possible to highly attenuated RSGs, especially above 20 \Msun,  by either interstellar or circumstellar dust. The latter is expected in a minority of RSGs, typically the most evolved ones, produced by pulsational mass-loss \citep[e.g.][]{Meynet2015}.
    \item  To enable some discussion about the statistics of older vs. younger RSGs, we define a sub-population of `young' RSGs, mostly with masses above 14 \Msun, with the thin dashed polygon region in Figure \ref{fig:rsgselection}. We refer to these RSGs in the discussion about spatial distribution in Section \ref{sec:spatialprops}. 
\end{enumerate} 
\begin{figure*}
    \centering
    \includegraphics[width=\textwidth]{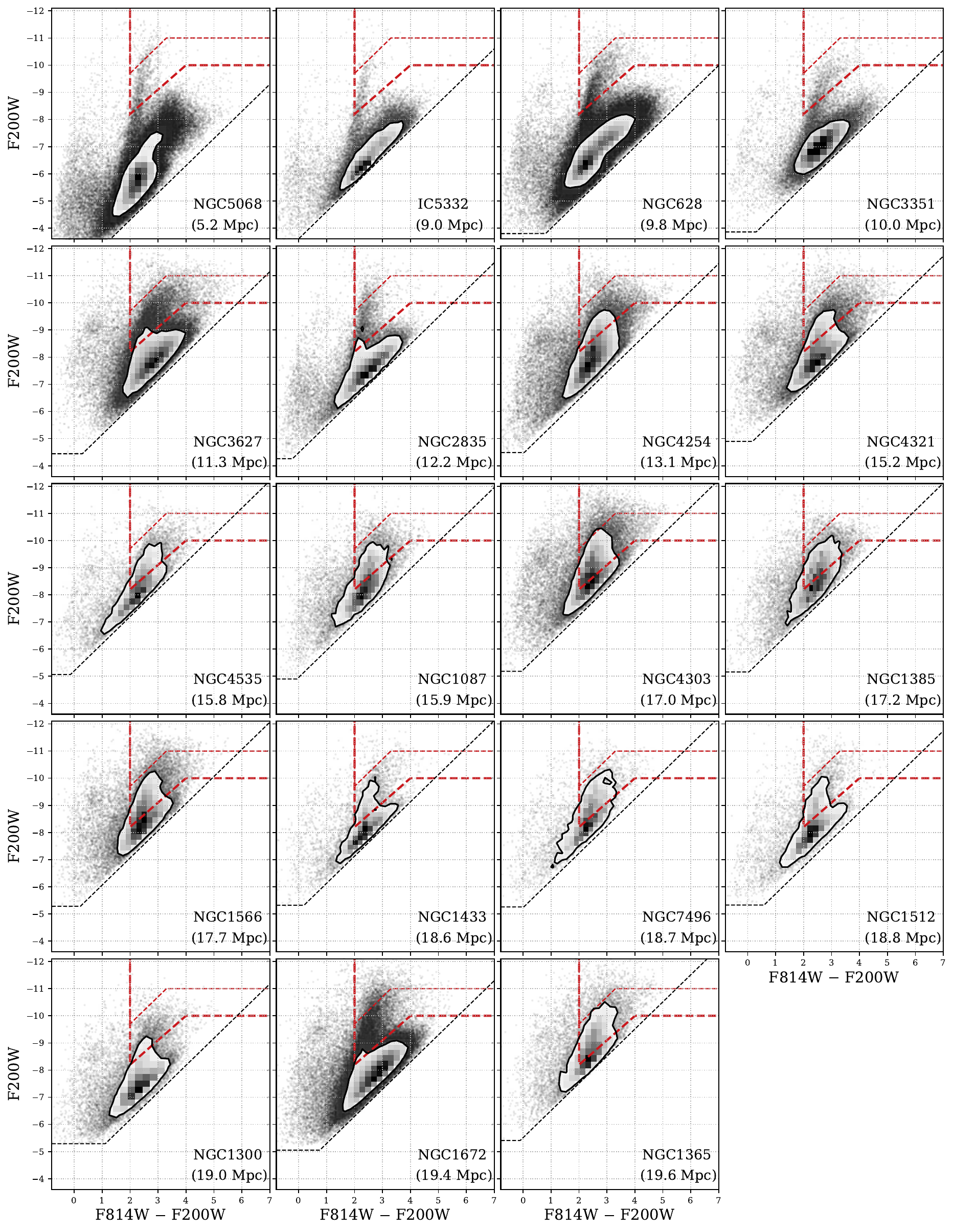}
    \caption{\ikcmd CMDs of all 19 galaxies, sorted by distance. Grey points show stars with magnitudes above the 5$\sigma$ \dolphot completeness limit (dashed black line) for F814W and F200W, as well as passing the crowding and sharpness cuts in Section \ref{sec:dolphot}. We also show the colored polygons from Figure \ref{fig:rsgselection}, showing the selection region for all RSGs in this paper (thick dashed) and younger RSGs (thin dashed).}
    \label{fig:rsgcmds}
\end{figure*}


\subsection{Comparison to data} \label{sec:compdata}
In the right panel of Figure \ref{fig:rsgselection}, we show a comparison of our RSG selection box with data from IC5332, where the RSG branch is clearly visible over the hill of AGB stars than span brightnesses all the way down to our completeness limit. IC5332 is our least extinguished target and is among our nearest, at a distance of $\sim$9 Mpc, so the RSGs are quite clearly visible. This validates the selection region we have defined in Section \ref{sec:rsgidentification}.

The CMDs of all 19 galaxies (consisting only of good stars from Section \ref{sec:dolphot}) along with the RSG selection box are shown in Figure \ref{fig:rsgcmds}. Our RSGs are $\sim$0.5-3.5 mag above the 5$\sigma$ DOLPHOT completeness limit for the 19 galaxies. In most nearby galaxies the RSG branch is clearly visible, while in others it is somewhat blurry, possibly due to a combination of differential extinction, photometric uncertainties, and metallicities. An upward protrusion along the wedge of the RSG selection box (Section \ref{sec:rsgidentification}) is still visible for all galaxies. The number of RSGs range between $<$10$^{3}$ in the lower mass, low SFR galaxies (e.g NGC 5068, IC5332) to $\gg$10$^3$ in higher mass, high SFR galaxies. Table \ref{tab:galaxies} reports the total number of RSGs and the number of younger RSGs found for each galaxy.

\begin{figure}[!htbp]
    \centering
    \includegraphics[width=\columnwidth]{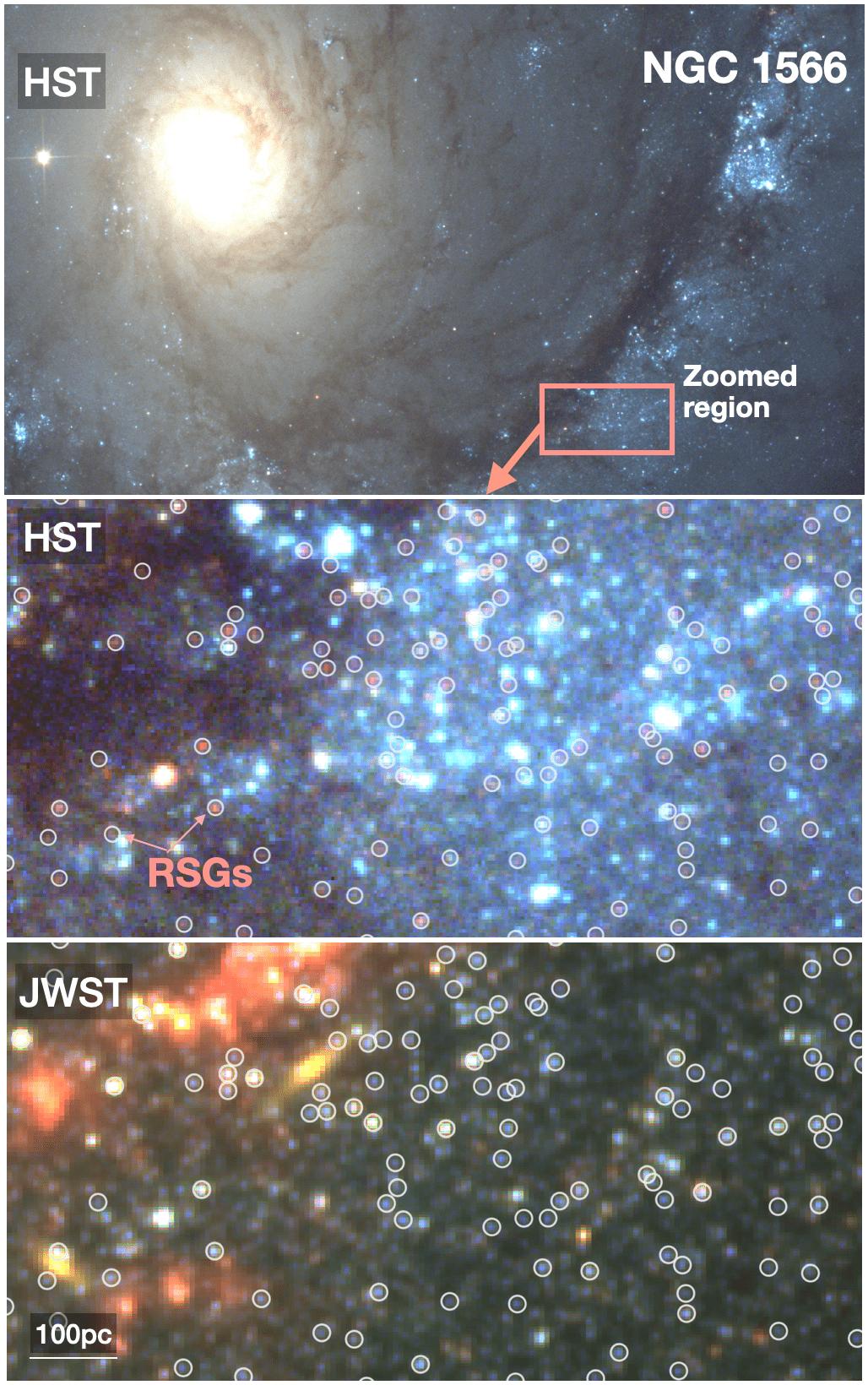}
    \caption{Zoom-in of a $\sim0.89\times0.47$~kpc$^2$ star-forming region in NGC 1566 at a distance of 17.7 Mpc showing the distribution of RSGs (white circles). Top and middle zoom-in panels show an \emph{rgb} HST image with $r$: F814W, $g$: F555W and $b$: F438W filters, and the bottom zoom-in panel is a JWST NIRCam \emph{rgb} image with $r$: F360M, $g$: F300M and $b$: F200W. The zoom-in region in the middle panel is shown in red in the top panel. Because RSGs typically peak around 1 \um, they appear as reddish point sources in the HST image (i.e.\ bright in F814W) and bluish in the NIRCam image (i.e.\ bright in F200W). Even at 17.7~Mpc, near the distance limit of our sample, these RSGs are clearly resolved in the HST and NIRCam images.}
    \label{fig:rsgzoomin}
\end{figure}

Figure~\ref{fig:rsgzoomin} shows a zoom-in region in the HST and NIRCam images for one of our distant targets, NGC 1566 at 17.7~Mpc. The figure visually confirms the recovery of point-like sources that satisfy the color-magnitude criteria for RSGs. Even in NGC 1566, which is near the distance limit of our sample (Table \ref{tab:galaxies}), the RSGs are clearly visible as point sources, appearing reddish in the HST image and bluish in the NIRCam image, due to their brightnesses peaking in the F814W to F200W filter range.

\subsection{Spatial Properties} \label{sec:spatialprops}

\begin{figure*}
    \centering
    \includegraphics[width=0.95\textwidth]{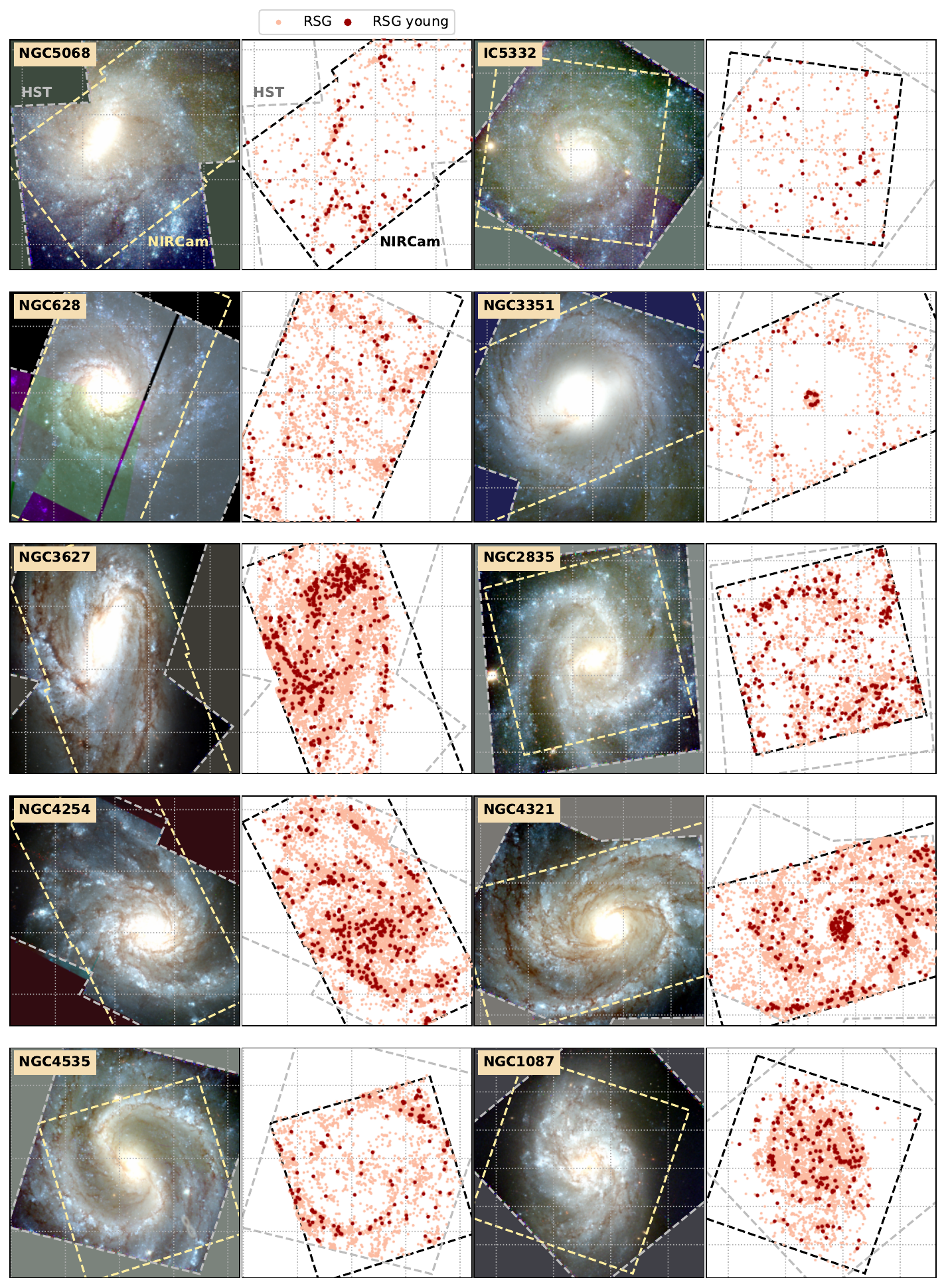}
    \caption{Spatial distribution of all RSGs (light-red), with the young RSGs from Figure \ref{fig:rsgselection} highlighted in darker red, shown alongside \emph{RGB} HST images, with \emph{B}: F438W/F435W, \emph{G}: F555W, and \emph{R}: F814W. Yellow/black dashed regions denote the NIRCam footprint, while gray dashed denote the HST footprints.}
    \label{fig:rsgspatial1}
\end{figure*}

\begin{figure*}[!htbp]
    \centering
    \includegraphics[width=0.95\textwidth]{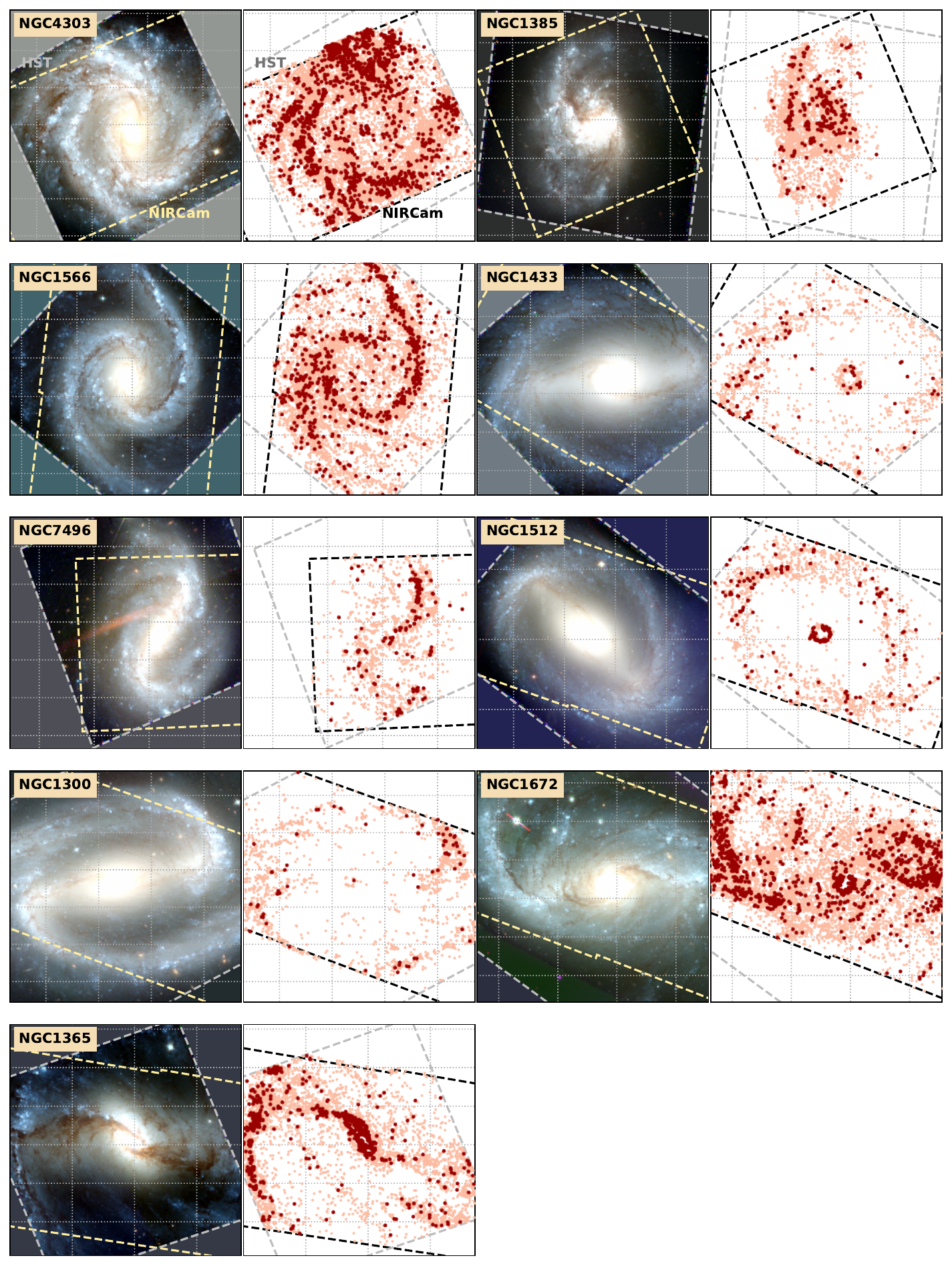}
    \caption{Same as Figure \ref{fig:rsgspatial1}, for the remaining galaxies in our sample.}
    \label{fig:rsgspatial2}
\end{figure*}

Figures \ref{fig:rsgspatial1} and \ref{fig:rsgspatial2} show the spatial distribution of the RSGs in all 19 galaxies, alongside the HST images for structural reference.\ The RSG distribution is clumpy, mostly appearing concentrated to the spiral arms, centers and other prominent areas of galaxies where recent star-formation is dominant. This is expected given the young ages ($\lesssim$40 Myr) of RSGs. Similar pattern RSGs being concentrated in star-forming regions is also observed in Local Group galaxies \citep[e.g][]{Massey2021}. For most galaxies, the RSGs appear concentrated in the spiral arms, especially in galaxies like NGC 1566, NGC 4321, and NGC 4254. In some cases like NGC 4321, NGC 1433, and NGC 1512, RSGs are also distributed along the central starburst rings. 
\begin{figure*}
    \centering
    \includegraphics[width=\textwidth]{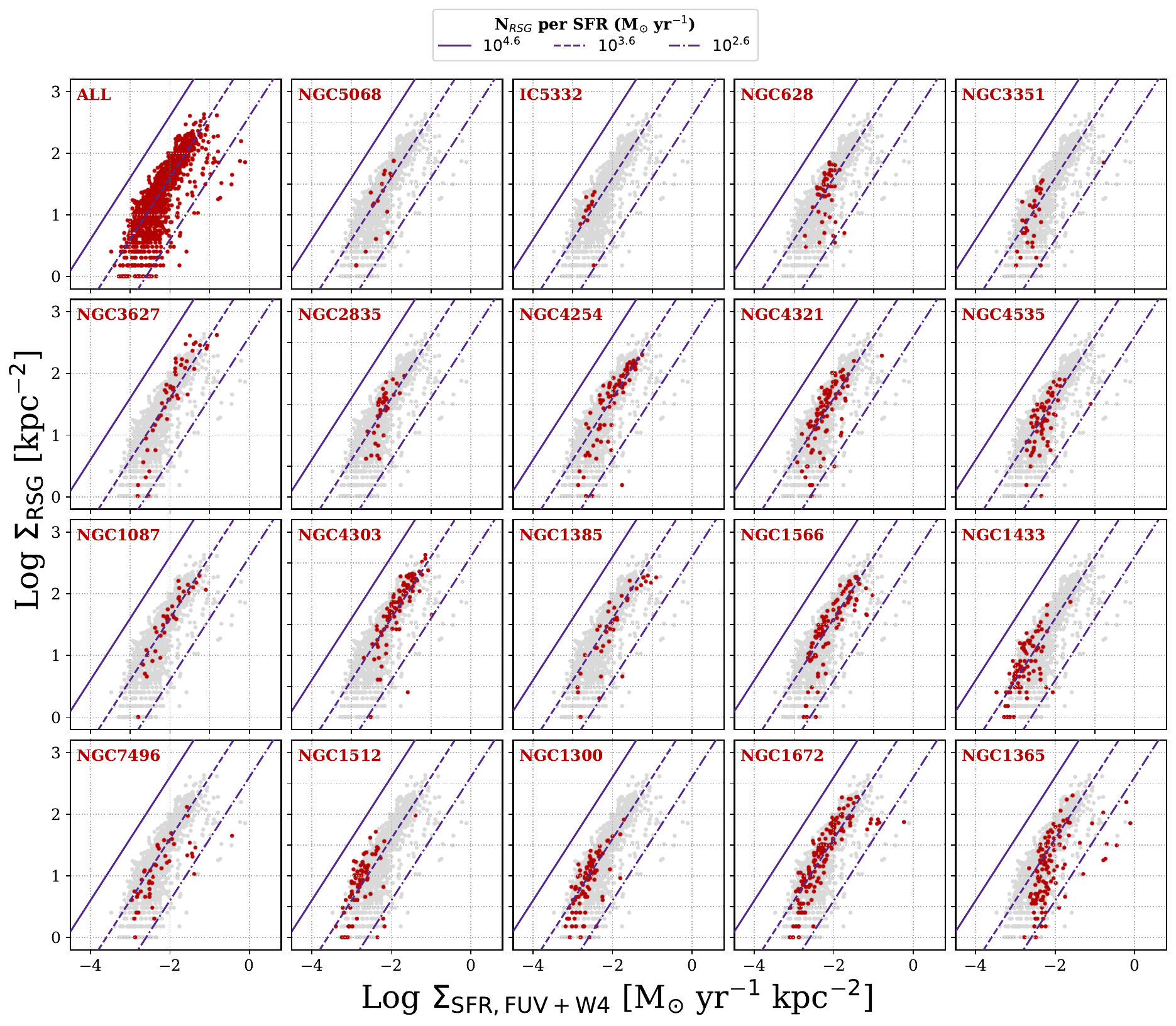}
    \caption{Correlation between number of RSGs and star-formation rate density (\Msunyrkpc) traced by GALEX FUV + WISE4 22$\mu$m, computed in 1.5 kpc diameter hexagons (Section \ref{sec:megatables}). The top left panel shows the combination of data from all galaxies in red. Subsequent panels show data for the individual galaxies in red, while the full sample (from panel 1) is shown in grey. Purple lines show constant number of RSGs per star-formation rate measured from FUV+W4 at 10$^{4.6}$, 10$^{3.6}$, and 10$^{2.6}$ RSGs per (M$_{\odot}$ yr$^{-1}$), corresponding to 1 RSG per 10$^{2.77}$, 10$^{3.77}$ and 10$^{4.77}$ \Msun of stars formed between 6-30 Myr (Section \ref{sec:results:sfrs}).}
    \label{fig:rsgfuv}
\end{figure*}
\begin{figure*}
    \centering
    \includegraphics[width=\textwidth]{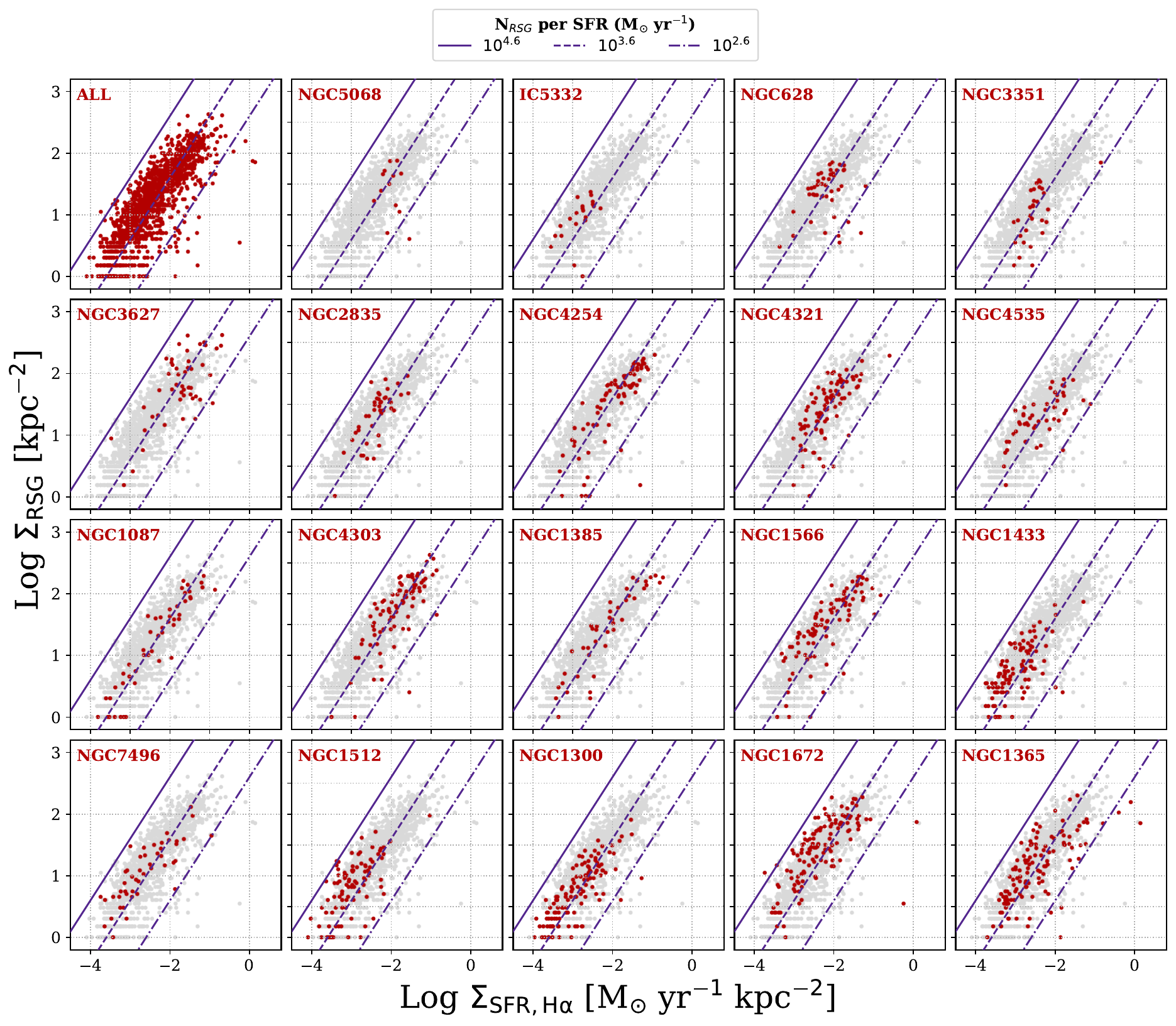}
    \caption{Same as Figure \ref{fig:rsgfuv}, but shown for SFRs from Balmer-decremented corrected H$\alpha$.}
    \label{fig:rsgha}
\end{figure*}

Figure \ref{fig:rsgspatial1} and \ref{fig:rsgspatial2} also illustrates difference between the more luminous and younger RSGs (darker red dots in Figure \ref{fig:rsgselection}, see Section \ref{sec:rsgidentification} for definition) and the bulk sample (lighter red dots). The younger RSGs appear more strongly correlated with the most active star-formation sites than the total RSG sample. We show this quantitatively based on the percentages of young vs.\ total RSGs in different environmental features in our galaxies, using masks defined in \cite{Querejeta2019}. We find that central starburst rings host a three times larger fraction of young RSGs compared to total RSGs. For example, NGC 3351 hosts 40\% of the younger RSG sample in the central ring compared to 13\% of the full RSG sample. Similarly the fraction in NGC 4321 jumps from 7\% for the total RSGs to 23\% for the younger RSGs. For NGC 1512, we find a jump of 10\% (total RSGs) to 29\% (younger RSGs). These fractions are broadly consistent with studies of the distribution recent star-formation versus environment using dust and photoionized gas emission by \cite{Querejeta2019} and \cite{Pathak2024}. In more detail, breaking the RSG sample into different mass brackets can potentially trace the stellar population age. Doing so will require careful estimates of the completeness in the presence of crowding, which will be explored in future work.

\section{Correlation with Local Star-Formation Rates} \label{sec:results:sfrs}

Figures~\ref{fig:rsgfuv} and \ref{fig:rsgha} show how the surface density of RSGs vary with the local stellar properties, specifically star-formation rates traced by FUV+W4 and \ha.

The RSG number density in Figure \ref{fig:rsgfuv} shows a strong correlation with the FUV+W4 star-formation rate surface density (\sigmasfr) over nearly two orders of magnitude between $\sim$10$^{-3}$ to 10$^{-1}$ \Msunyrkpc for the combined sample of 19 galaxies, with a Spearman rank correlation coefficient ($r_s$) of 0.82. We also found similarly strong correlations of $r_s=0.82$ with Balmer-decrement-corrected MUSE \ha SFR (Figure \ref{fig:rsgha}), and $r_s=0.75$ with WFI \ha+W4 based SFR from the PHANGS MegaTables (Section \ref{sec:obs}). In contrast, the RSG number densities show weaker correlations with the total stellar mass, traced by 3.4\mum and 3.6\mum emission, with an $r_s=0.56$. At a given SFR, the number density of RSGs varies by about 0.5~dex, bottoming out to 1--2 RSGs per kpc$^{2}$ at SFRs $<10^{-3}$~\Msunyrkpc. 

These correlations are consistent with the RSGs being young ($>$10~\Msun, ages $\lesssim$30 Myr) stars such that their abundance reflects star formation activity on similar timescales to commonly used star formation tracers
\citep[see reviews in][]{Kennicutt1998, Kennicutt2012, Schinnerer2024}. FUV emission traces star-formation in the last $\sim100$~Myr, with a simple stellar population emitting roughly equal total energy between $0-10$~Myr and $10-100$~Myr. The light-weighted \ha traces star-formation in the last 10~Myr, with a simple stellar population emitting half of its total luminosity before 3 Myr. Mid-IR emission, including the 22$\mu$m emission included in our FUV+WISE4 tracer, has a more ambiguous timescale but is often taken to reflect star formation over a similar timescale to the FUV. Meanwhile, the total stellar mass is mostly dominated by the older ($\gtrsim$Gyr) stellar population which does not produce RSGs. 

We also estimate the number of RSGs per mass of recently formed stars based on the data in Figure \ref{fig:rsgfuv}. The number of RSGs per FUV+W4 SFR is about 10$^{3.60\pm0.27}$ per (M$_{\odot}$ yr$^{-1}$), excluding bins with low RSG number counts ($<$3 RSGs per kpc$^2$) and high SFRs where extinction and crowding effects could be high ($>$10$^{-1.2}$ \Msunyrkpc). Repeating the same with the \ha based SFRs in Figure \ref{fig:rsgha}, we obtain 10$^{3.70\pm0.35}$ RSGs per (\Msunyr), slightly higher but within error bars of the FUV+W4 based SFRs.  If we assume constant star-formation histories over the last 100 Myr, and multiply the FUV+W4 SFR with the typically assumed duration of RSG formation above 10 \Msun ($\sim$6-30 Myr), we obtain the \emph{production rate}, i.e the number of RSGs per \Msun formed, to be 10$^{-3.77\pm0.27}$ RSGs per \Msun for the full 19 galaxy sample. These are shown as purple lines in Figure \ref{fig:rsgfuv}. Visually, the production rate appears to be consistent across the 19 galaxy sample. Some deviation from these values can be seen at high SFRs ($\gtrsim$10$^{-1.2}$ \Msunyrkpc) in NGC 1672 and NGC 1365, as well as at SFRs $\lesssim$10$^{-3}$ \Msunyrkpc where uncertainties due to low number statistics for RSGs, and completeness limit of \emph{WISE W4} can be factors.

We checked these observed ratios against synthetic populations generated with PARSEC to perform a sanity check. Using the \texttt{CMD} interface, we created 20 realizations of synthetic composite stellar populations at solar metallicity and zero extinction, each consisting of simple stellar populations with initial masses of $10^4$~\Msun, formed at ages between 3 and 100~Myr at an interval of $\Delta t=2$~Myr. This is equivalent to generating a composite stellar population at a constant star-formation rate of $5\times10^{-3}$~\Msunyr, which falls within the range of SFRs observed over kpc-scales in Figure \ref{fig:rsgfuv}. We identify RSGs as the stars with $\log(T/K)<3.6$ and $\log(L/L_{\odot})\geq4.3$, which covers approximately the same \ikcmd for the RSGs in Figure \ref{fig:rsgselection}. Averaged over the 20 realizations, we get $40\pm7$ RSGs. When we divide this by the modeled SFR, we get roughly $10^{3.9\pm0.8}$ RSGs per (M$_\odot$ yr$^{-1}$). When divided by the mass of stars formed at ages between 6 and 30~Myr, we derive a ratio of 10$^{-3.50\pm0.08}$ RSGs per \Msun. This is slightly higher than our observed ratio of $10^{3.6}$ RSGs per (M$_\odot$ yr$^{-1}$) and 10$^{-3.77\pm0.27}$ RSGs per \Msun\ formed between 6 and 30 Myr.

Thus our modeling predicts approximately twice as many RSGs as we recover given the estimated star formation rates. The discrepancy could reflect incompleteness due to crowding or extinction, particularly in the lower mass, lower luminosity RSGs, which are also more numerous. Alternatively, it could reflect a systematic overestimate of the SFR or systematic variations in the star formation history, e.g., gaps on $\approx 6{-}30$~Myr timescales, across our sample. A significant overestimate of the SFR seems unlikely given the good correspondence between H$\alpha$+H$\beta$ based SFRs and FUV+WISE4 SFRs shown by \citet{Belfiore2023} specifically for this sample. And we would expect variations in the SFH on short timescales to produce large fluctuations in Fig. \ref{fig:rsgfuv}. A combination of the above could e.g explain the notable low deviates in NGC 1365 or NGC 4321 in Fig. \ref{fig:rsgfuv} at $\gtrsim$10$^{-1.2}$ \Msunyrkpc. Alternatively, physical reasons related to stellar populations such as evolution of a significant number of RSGs in interacting binary systems, could also lead to a reduction in RSGs per mass of stars \citep[e.g][]{Dorn-Wallenstein2018, Dorn-Wallenstein2020}.




\section{Conclusion} \label{sec:conclusion}
We have recovered a catalog of 97057 RSGs---the largest single survey of red-supergiants (RSGs) ---in a sample of 19 galaxies with stellar masses between $10^9$ and $10^{11}$ \Msun from the PHANGS-HST and PHANGS-JWST surveys. We derived the catalog from \dolphot PSF photometry in the filters closest to the peak wavelength of RSGs---UVIS/WFC3 F814W and NIRCAM F200W bands. RSGs were selected from the predicted positions of stars cooler than 4000 K and above $>$10 \Msun in \ikcmd, as predicted by stellar isochrone models from PARSEC. The region was designed to minimize contamination with AGB stars, foreground dwarfs, and hotter supergiants.

We mainly carry out a zeroth order analysis of the RSG population in this paper, focusing on the number and spatial statistics, as well as correlation with kpc-scale measurements of star-formation rates from the PHANGS MegaTables from \cite{megatable}. We find that:
\begin{itemize}
\itemsep -1pt
\item The number of RSGs recovered in these galaxies ranges from a few hundred (e.g 453 in IC5332) to over 10$^4$ (NGC~4303). 
\item The distribution of the RSGs closely trace prominent sites of recent star-formation in galaxies, such as central starburst rings, and spiral arms. Younger ($>$14 \Msun) RSGs are more strongly spatially correlated with high star-formation sites, such as central starburst rings, than the total RSG sample.

\item The RSG number density is strongly correlated with the FUV+W4 based star-formation rates, with a Spearman correlation coefficient $r_s$=0.82. Similarly strong correlation is obtained with SFRs based on \ha, and only a weak correlation ($r_s$=0.57) is obtained with stellar mass.

\item We find a consistent production rate of 10$^{3.6}\pm0.27$ RSG per (\Msunyr), or 1 RSG per 10$^{3.77\pm0.27}$ \Msun of stars formed between 6-30 Myr in the 19 galaxies, assuming the average star-formation rate during this period given by FUV+W4 based SFRs. The observed rates are slightly lower than predictions of 1 RSG per 10$^{3.5\pm0.08}$ \Msun from stellar population models, and could be due to assumptions about star-formation histories, incompleteness in high extinction/crowding regions, and physical factors such as binary evolution. 
\end{itemize}

The paper breaks new ground by pushing systematic searches of RSGs in a large sample of galaxies with distances up to 20~Mpc by leveraging the combined power of HST and JWST NIRCam, a feat that was impossible with previous observatories. Our catalog will be published with the journal, and serve as a valuable resource for characterizing progenitors of future Type II supernovae in the PHANGS sample \citep[e.g][]{vandyk2023, Kilpatrick2025}, age-dating and obtaining census of star-formation in the 6-30 Myr range \citep[e.g][]{Eldridge2020,Hannon2022}, and studying correlations of future supernova sites with star clusters, associations and molecular clouds \citep[e.g][]{Kangas2017,Sarbadhicary2023,MaykerChen2023,MaykenChen2024}. 

In upcoming papers, we will continue refining the catalog with artificial star tests for more systematically characterizing the completeness in different regions of the galaxies, and estimating bolometric properties and line-of-sight extinction \citep{Gordon2016,Lindberg2025}. Our catalog can be cross-matched with MIRI data for investigation on dust production in RSGs,  expanding upon the works by \cite{hassani2025} and \cite{Maschmann2025}. Future observations using JWST filters near 1 \mum, such as F090W and F115W can significantly improve the RSG selection at lower masses ($\lesssim$10 \Msun) by not only being higher-resolution blue filters than F814W, but also less sensitive to extinction.

\begin{acknowledgments}
SKS and DT acknowledge support from HST grants GO 17502 and AR 17572. SKS acknowledges helpful discussions with the PHANGS Star Clusters group at STSCI. This work is based on observations made with the NASA/ESA/CSA James Webb Space Telescope (program \#2107) and the NASA/ESA Hubble Space Telescope (program \#15654).   
The data were obtained from the Mikulski Archive for Space Telescopes at the Space Telescope Science Institute, which is operated by the Association of Universities for Research in Astronomy, Inc., under NASA contract NAS 5-03127 for JWST and 5-26555 for HST.
The specific observations analyzed can be accessed via \dataset[DOI: 10.17909/t9-r08f-dq31]{https://doi.org/10.17909/t9-r08f-dq31}, and \dataset[DOI: DOI/ew88-jt15]{https://archive.stsci.edu/doi/resolve/resolve.html?doi=10.17909/ew88-jt15}.

M.B. acknowledges support by the ANID BASAL project FB210003. This work was supported by the French government through the France 2030 investment plan managed by the National Research Agency (ANR), as part of the Initiative of Excellence of Université Côte d’Azur under reference No. ANR-15-IDEX-01. This research was funded, in whole or in part, by the French National Research Agency (ANR), grant ANR-24-CE92-0044 (project STARCLUSTERS).  We thank the German Science Foundation DFG for financial support in project STARCLUSTERS (funding ID KL 1358/22-1 and SCHI 536/13-1).
HAP acknowledges support from the National Science and Technology Council of Taiwan under grant 113-2112-M-032-014-MY3. R.S.K. and S.O.G. acknowledges financial support from the ERC via Synergy Grant ``ECOGAL'' (project ID 855130),  from the German Excellence Strategy via the Heidelberg Cluster ``STRUCTURES'' (EXC 2181 - 390900948), from the German Ministry for Economic Affairs and Climate Action in project ``MAINN'' (funding ID 50OO2206), and from DFG and ANR for project ``STARCLUSTERS'' (funding ID KL 1358/22-1). 
\end{acknowledgments}

\begin{contribution}

SKS led the analysis in this paper in close collaboration with DT, AKL, and JCL. LU provided much of the framework for DOLPHOT analysis. AKL and JS are among the architects of the PHANGS MegaTable, and helped with analyses. All authors contributed to the overall science interpretations and writing.

\end{contribution}

\facilities{HST(WFC3, ACS), JWST (NIRCam)}

\software{\texttt{astropy} \citep{Astropy}, \texttt{matplotlib} \citep{matplotlib}, \texttt{scipy} \citep{scipy}, \texttt{numpy} \citep{numpy}, \texttt{regions} \citep{regions}, \texttt{CARTA} \citep{carta1,carta2}, \dolphot \citep{Dolphin2000}, \texttt{corner.py} \citep{corner}
          }

\bibliography{main}{}
\bibliographystyle{aasjournalv7}

\end{document}

%% file: authors.tex

\newcommand{\JHU}{\affiliation{Department of Physics and Astronomy, The Johns Hopkins University, Baltimore, MD 21218 USA}}

\newcommand{\OSUCCAPP}{\affiliation{Center for Cosmology Astro-Particle Physics, The Ohio State University, Columbus, Ohio 43210, USA}}

\newcommand{\OSUAstro}{\affiliation{Department of Astronomy, The Ohio State University, Columbus, Ohio 43210, USA}}
\newcommand{\UniCA}{\affiliation{Université Côte d'Azur, Observatoire de la Côte d'Azur, CNRS, Laboratoire Lagrange, 06000, Nice, France}}
\newcommand{\UWyoming}{\affiliation{Department of Physics and Astronomy, University of Wyoming, Laramie, WY 82071, USA}}

\newcommand{\STScI}
{\affiliation{Space Telescope Science Institute, 3700 San Martin Drive, Baltimore, MD 21218, USA}}

\newcommand{\ESO}{\affiliation{European Southern Observatory, Karl-Schwarzschild Stra{\ss}e 2, D-85748 Garching bei M\"{u}nchen, Germany}}

\newcommand{\UoA}{\affiliation{Department of Physics, University of Arkansas, Fayetteville, AR 72701, USA}}

\newcommand{\JBCA}{\affiliation{UK ALMA Regional Centre Node, Jodrell Bank Centre for Astrophysics, Department of Physics and Astronomy, The University of Manchester, Oxford Road, Manchester M13 9PL, UK}}

\newcommand{\TKU}{\affiliation{Department of Physics, Tamkang University, No.151, Yingzhuan Road, Tamsui District, New Taipei City 251301, Taiwan}}

\newcommand{\UCT}{\affiliation{Department of Astronomy, University of Cape Town, Rondebosch, Cape Town, 7700, South Africa}}

\newcommand{\UKy}{\affiliation{Department of Physics and Astronomy, University of Kentucky, 506 Library Drive, Lexington, KY 40506, USA}}

\newcommand{\ITA}{\affiliation{Universit\"{a}t Heidelberg, Zentrum f\"{u}r Astronomie, Institut f\"{u}r Theoretische Astrophysik, Albert-Ueberle-Str.\ 2, 69120 Heidelberg, Germany}}
\newcommand{\IWR}{\affiliation{Universit\"{a}t Heidelberg, Interdisziplin\"{a}res Zentrum f\"{u}r Wissenschaftliches Rechnen, Im Neuenheimer Feld 225, 69120 Heidelberg, Germany}}

\newcommand{\UNAM}{\affiliation{Instituto de Astronom\'ia, Universidad Nacional Aut\'onoma de M\'exico, Unidad Acad\'emica en Ensenada, Km 103 Carr. Tijuana$-$Ensenada, Ensenada, B.C.,\\C.P. 22860, M\'exico}}

\newcommand{\STScIESA}{\affiliation{AURA for the European Space Agency (ESA), Space Telescope Science Institute, 3700 San Martin Drive, Baltimore, MD 21218, USA}}

\author[0000-0002-6313-4597]{Sumit K. Sarbadhicary}
\JHU
\email[show]{ssarbad1@jh.edu}

\author[0000-0002-8528-7340]{David Thilker}
\JHU
\email{dthilker@jhu.edu}

\author[0000-0002-2545-1700]{Adam K. Leroy}
\OSUAstro
\OSUCCAPP
\email{leroy.42@osu.edu}

\author[0000-0003-0946-6176]{Janice C. Lee}
\STScI
\email{jlee@stsci.edu}

\author[0000-0002-8553-1964]{Amirnezam Amiri}
\UoA
\email{amirnezamamiri@gmail.com}

\author[0000-0002-5259-2314]{Gagandeep S. Anand}
\STScI
\email{ganand@stsci.edu}  
\author[0000-0003-0410-4504]{Ashley.~T.~Barnes}
\ESO
\email{ashley.barnes@eso.org}

\author[0000-0003-0946-6176]{Médéric Boquien}
\UniCA
\email{mederic.boquien@oca.eu}

\author[0000-0002-5782-9093]{Daniel~A.~Dale}
\UWyoming
\email{DDale@uwyo.edu}
\author[0000-0002-2885-6172]{Simthembile Dlamini}
\UCT
\email{simther4111@gmail.com}

\author[0000-0001-6708-1317]{Simon~C.~O.\ Glover}
\ITA
\email{glover@uni-heidelberg.de}

\author[0000-0002-0560-3172]{Ralf S.\ Klessen}
\ITA,\IWR
\email{klessen@uni-heidelberg.de}

\author[0000-0003-3917-6460]{Kirsten L. Larson}
\STScIESA
\email{kilarson@stsci.edu}

\author[0000-0001-6038-9511]{Daniel Maschmann}
\UWyoming
\email{dmaschma@uwyo.edu}
\author[0000-0002-1370-6964]{Hsi-An Pan}
\TKU
\email{hapan@gms.tku.edu.tw}

\author[0000-0003-0378-4667]{Jiayi Sun}
\UKy
\email{jiayi.sun@uky.edu}

\author[0000-0001-7130-2880]{Leonardo \'Ubeda}
\STScI
\email{lubeda@stsci.edu}

\author[0000-0002-0012-2142]{Thomas G. Williams}
\JBCA
\email{thomas.g.williams@manchester.ac.uk}

\author[0000-0001-8289-3428]{Aida Wofford}
\UNAM
\email{awofford@astro.unam.mx}

\collaboration{all}{PHANGS Collaboration}